\documentclass[10pt,twoside]{hsqcd}
\usepackage{epsf,epsfig,amsmath,amssymb,bm}
\usepackage{graphicx}
\usepackage{amssymb}

\newcommand{\vecc}[1]{\mbox{\boldmath $#1$}}

\begin{document}

\title{{\hfill RUB-TPII-06/08}\\ [1cm]ANOMALOUS DIMENSIONS OF 
       TRANSVERSE-MOMENTUM DEPENDENT PARTON DISTRIBUTION 
       FUNCTIONS\thanks{Invited talk presented by the first author at International 
        Workshop ``Hadron Structure and QCD'' (HSQCD'2008), 
        June 30 - July 4, 2008, Gatchina, Russia}}
\author{\underline{N.~G.~STEFANIS}$^{1,a}$
               and I.~O.~CHEREDNIKOV$^{2,b}
\thanks{The work of I.O.C was supported by the Alexander
        von Humboldt Stiftung.}$     \\ \\
$^1$Institut f\"ur Theoretische Physik II, \\
    Ruhr-Universit\"at Bochum, \\
    D-44780 Bochum, Germany    \\
$^a$Email: stefanis@tp2.ruhr-uni-bochum.de\\
$^2$Bogoliubov Laboratory of Theoretical Physics, JINR, \\
    141980 Dubna, Russia       \\
$^b$Email: igorch@theor.jinr.ru }

\maketitle

\begin{abstract}
\noindent We discuss recent developments in the understanding of
gauge-invariant transverse-momentum dependent (TMD) parton-distribution
functions (PDF).
We compute the leading-order $\overline{\text{MS}\vphantom{^1}}$-scheme
anomalous dimension of such a quantity in the light-cone gauge and show
that it receives a contribution that can be associated with a cusp
obstruction at transverse light-cone infinity.
This anomalous dimension is intimately related to expectation values
composed of fields and eikonal factors along a cusped contour and is
absent in covariant gauges.
The implications of these findings are addressed and a modified
definition of TMD PDFs is proposed.
\end{abstract}



\markboth{\large \sl \underline{N.~G.~STEFANIS} \& I.~O.~CHEREDNIKOV
\hspace*{2cm} HSQCD 2008} {\large \sl \hspace*{1cm} ANOMALOUS DIMENSIONS
                                                    OF TMD PDFs}

\section{Introduction}
In this report, recent theoretical results on TMD PDFs are presented,
giving particular emphasis on their renormalization-group properties
in conjunction with local color-gauge invariance.
The latter is ensured by the presence of Wilson lines (gauge links)
that are required to obtain gauge-invariant definitions of PDFs,
integrated over the parton transverse momentum, and also
unintegrated, i.e., TMD PDFs (see, for reviews \cite{CSS89,ER80}).
These quantities encapsulate the nonperturbative parts of the QCD
processes, being themselves universal, with a (large) momentum-scale
dependence controlled by perturbative QCD in the form of
renormalization-group type evolution equations, like the
Dokshitzer--Gribov--Lipatov--Altarelli--Parisi) (DGLAP) evolution
equation.

Central to the solution of these equations is the knowledge of the
appropriate anomalous dimensions, calculable within perturbative
QCD.
One of the main purposes of this presentation is to discuss the
physical impact of the leading-order anomalous dimensions in the
light-cone gauge of matrix elements containing local operators and
path-ordered exponentials along gauge contours that comprise
transverse segments extending to light-cone infinity.
The particular relevance of the anomalous dimensions originates from
the fact that, in contrast to the gauge links which are path-dependent,
they are local objects and ensure the gauge invariance of
contour-dependent operators in terms of ``logarithmic'', i.e.,
additive, Ward-Takahashi/Slavnov-Taylor identities.
Because the anomalous dimensions of such operators ensue from
obstructions, like endpoints, cusps, or self-intersections, one is able
to analyze the renormalization-group properties of TMD PDFs solely by
means of these quantities without any reference to the explicit, in
general complicated, gauge-link structure.
It was shown in \cite{BJY02} (see also \cite{BHS02,JY02,BMP03}) that
the inclusion of transverse gauge links at light-cone infinity is
indispensable for the restoration of full gauge invariance of the
TMD PDFs in the light-cone gauge.
Otherwise the gauge freedom in the light-cone gauge is not completely
exhausted and, therefore, the light cone gauge is insufficient to
trivialize the interaction of the struck quark with the gluon field of
the spectators.
Moreover, the transverse gauge link is responsible for the final
(initial) state interactions.

\section{Gauge-contour dependent TMD PDFs}
\label{sec:contour-dep-PDF}

We begin with the operator definition of the TMD distribution of a
quark with momentum
$k_\mu = (k^+, k^-, \vecc k_\perp)$
in a quark with momentum $p_\mu = (p^+, p^-, \vecc 0_\perp)$, with
decomposed gauge contours going through light-cone infinity:
\begin{equation}
\begin{split}
   f_{q/q}(x, \mbox{\boldmath$k_\perp$})
 ={} &
   \frac{1}{2} \!
   \int \! \frac{d\xi^- d^2
   \mbox{\boldmath$\xi_\perp$}}{2\pi (2\pi)^2}\,
   \exp\left(- i k^+ \xi^- \!
   +\!  i \mbox{\boldmath$k_\perp$}
\cdot \mbox{\boldmath$\xi_\perp$}\right)
   \left\langle  q(p) |\bar \psi (\xi^-, \vecc \xi_\perp)
   [\xi^-, \mbox{\boldmath$\xi_\perp$};
   \infty^-, \mbox{\boldmath$\xi_\perp$}]^\dagger \right.\\
& \times [\infty^-, \mbox{\boldmath$\xi_\perp$};
   \left. \infty^-, \mbox{\boldmath$\infty_\perp$}]^\dagger
   \gamma^+[\infty^-, \mbox{\boldmath$\infty_\perp$};
   \infty^-, \mbox{\boldmath$0_\perp$}]
   [\infty^-, \mbox{\boldmath$0_\perp$};0^-, \mbox{\boldmath$0_\perp$}] \right. \\
& \times \left. \psi (0^-,\mbox{\boldmath$0_\perp$}) |q(p)\right\rangle \
   |_{\xi^+ =0}\ . \vspace{-1cm}
\label{eq:TMD-PDF}
\end{split}
\end{equation}
Here the gauge links, lightlike and transverse, respectively, are
defined by the following path-ordered exponentials
\begin{equation}
\begin{split}
[ \infty^-, \mbox{\boldmath$z_\perp$}; z^-, \mbox{\boldmath$z_\perp$}]
\equiv {} &
 {\cal P} \exp \left[
                     i g \int_0^\infty d\tau \ n_{\mu}^- \
                      A_{a}^{\mu}t^{a} (z + n^- \tau)
               \right] \\
[ \infty^-, \mbox{\boldmath$\infty_\perp$};
 \infty^-, \mbox{\boldmath$\xi_\perp$}]
\equiv {} &
 {\cal P} \exp \left[
                     i g \int_0^\infty d\tau \ \mbox{\boldmath$l$}
                     \cdot \mbox{\boldmath$A$}_{a} t^{a}
                     (\mbox{\boldmath$\xi_\perp$}
                     + \mbox{\boldmath$l$}\tau)
               \right] \, ,
\label{eq:gauge-links}
\end{split}
\end{equation}
where the two-dimensional vector $\vecc l$ is arbitrary with no
influence on the (local) anomalous dimensions we are interested in.

\begin{figure}[t]
\centering
\includegraphics[scale=0.5,angle=90]{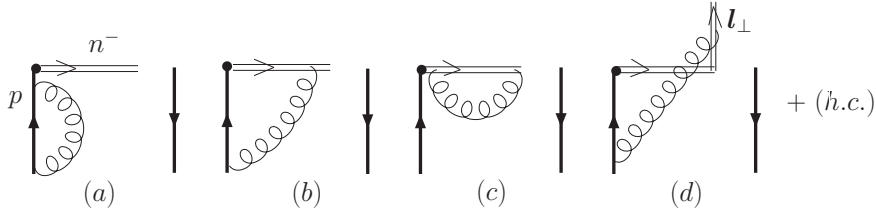}~~
\caption{One-loop radiative corrections (curly lines) contributing
         UV-divergences to the TMD PDF in a general covariant gauge,
         with double lines denoting gauge links.
         Diagrams (b) and (c) are absent in the light-cone gauge, while
         the Hermitian conjugate (``mirror'') diagrams (not shown)
         are abbreviated by $(h.c)$.
\label{fig:se_gluon}}
\end{figure}

We have calculated \cite{CS07} the one-gluon exchange contributions to
the (unpolarized) TMD PDF of a quark in a quark and identified their UV
divergences (see Fig.\ \ref{fig:se_gluon}).
It turns out that in the light-cone gauge, those contributions stemming
from the interactions with the gluon field of the transverse gauge link
cancel all terms that bear a dependence on the pole prescription
applied to go around the light-cone singularities of the gluon
propagator.
As such, we have employed the retarded, advanced, and principal-value
prescriptions (see for details \cite{CS07}).
Taking into account the Hermitian conjugate contributions, we have
identified an UV-divergent contribution that is absent in covariant
gauges (e.g., the Feynman gauge) and can be conceived as originating
from a non-trivial cusp-like junction point of the individual
(transverse) gauge contours in the TMD PDF.
Referring for details to \cite{CS07}, we here display only
the complete UV-divergent part of the TMD PDF:
\begin{eqnarray}
  {\Sigma}^{(a+d)}_{\rm UV} (p, \mu, \alpha_s ; \epsilon)
& \!\!\!\!\! = \!\!\!\!\! &
  - \frac{\alpha_s}{\pi}\ C_{\rm F} \frac{1}{\epsilon}
    \left[\frac{1}{4}- \frac{ \gamma^+ \hat p}{2 p^+}
  \left( 1 +  \ln \frac{\eta}{p^+} - \frac{i\pi}{2}
  - i  \pi \ C_\infty + i \pi C_\infty \right)
    \right]
\nonumber \\
& \!\!\!\!\! = \!\!\!\!\! &
  - \frac{\alpha_s}{\pi}\ C_{\rm F}\   \frac{1}{\epsilon}
    \left[1 - \frac{ \gamma^+ \hat p}{2 p^+}
    \left( 1  + \ln \frac{\eta}{p^+}
    - \frac{i\pi}{2} \right) \right] \ .
\label{eq:UV-parts}
\end{eqnarray}
To complete the argument, take into account that
$
  \frac{ \gamma^+ \hat p \gamma^+}{2 p^+} = \gamma^+
$
and recall that we have to include the \emph{mirror} {\it h.c.}
counterparts of the evaluated diagrams.
These yield complex-conjugated contributions, so that the imaginary
terms mutually cancel.
Hence, the UV-divergent part of diagrams (a) and (d) contains only
contributions due to the $p^+$-dependent term:
\begin{equation}
  \Sigma_{\rm UV}^{\rm (a+d)}(\alpha_s, \epsilon)
=
   2\frac{\alpha_s}{\pi}C_{\rm F} \left[ \frac{1}{\epsilon}
   \left( \frac{3}{4}
  + \ln \frac{\eta}{p^+} \right) - \gamma_E + \ln 4\pi \right] \, .
\label{eq:Sigma-UV-part}
\end{equation}
There is an extra anomalous dimension associated with the
$p^+$-dependent term which at the one-loop level reads
$
 \left(\gamma
=
  \frac{\mu}{2}
  \frac{1}{Z}
  \frac{\partial\alpha_s}{\partial\mu}
  \frac{\partial Z}{\partial\alpha_s}
\right)
$
\begin{equation}
  \gamma_{\rm 1-loop}^{\rm LC}
=
  \frac{\alpha_s}{\pi}C_{\rm F}\left(\frac{3}{4}
  + \ln \frac{\eta}{p^+} \right)
=
  \gamma_{\rm smooth} - \delta \gamma \, .
\label{eq:extra-dim}
\end{equation}

The renormalization effect due to the gluon corrections on the
cusped junction point of the contours gives rise to the
anomalous-dimension defect $\delta \gamma$ that has to be compensated
in order to recover the same expression as in a covariant gauge
according to the factorization proof.
Moreover, it entails a modification in the multiplication rule for
gauge links, according to \cite{CS07}
\begin{equation}
  \gamma_{\mathcal{C}}
=
  \gamma_{\mathcal{C}_{1}^{\infty} \cup\, \mathcal{C}_{2}^{\infty}
           }
  +\gamma_{\rm cusp}
  ~~ \Longleftrightarrow ~~
  [2,1|\mathcal{C}]
=
  [2,\infty|\mathcal{C}_{2}^{\infty}]^{\dag}
  [\infty,1|\mathcal{C}_{1}^{\infty}]
  {\rm e}^{i \Phi_{\rm cusp}
          } \, .
\label{eq:mod-mult-rule}
\end{equation}
The crucial point here is to understand that the defect of the
anomalous dimension can be identified with the universal cusp anomalous
dimension \cite{KR87}, i.e.,
\begin{equation}
\begin{split}
   & \gamma_{\rm cusp} (\alpha_s, \chi)
= \frac{\alpha_s}{\pi}C_{\rm F} \ (\chi \coth \chi - 1 ) \ , \\
& \frac{d}{d \ln p^+} \ \delta \gamma
= \lim_{\chi \to \infty}
  \frac{d}{d \chi} \gamma_{\rm cusp} (\alpha_s, \chi)
= \frac{\alpha_s}{\pi}C_{\rm F} \ .
\label{eq:cusp-an-dim}
\end{split}
\end{equation}
Then, to dispense with the anomalous-dimension artefact, the original
TMD PDF will be redefined in terms of extra eikonal factors akin to the
soft counter terms of Collins and Hautmann \cite{CH00}:
\begin{equation}
  R
\equiv
 \Phi (p^+, n^- | 0) \Phi^\dagger (p^+, n^- | \xi)
\label{eq:R}
\end{equation}
with eikonal factors given by
\begin{eqnarray}
\Phi (p^+, n^- | 0 )
 & = &
  \left\langle 0
  \left| {\cal P}
  \exp\Big[ig \int_{\mathcal{C}_{\rm cusp}}\! d\zeta^\mu
           \ t^a A^a_\mu (\zeta)
      \Big]
  \right|0
  \right\rangle \
\ , \\
  \Phi^\dagger (p^+, n^- | \xi )
 & = &
  \left\langle 0
  \left| {\cal P}
  \exp\Big[- ig \int_{\mathcal{C}_{\rm cusp}}\! d\zeta^\mu
           \ t^a A^a_\mu (\xi + \zeta)
      \Big]
  \right|0
  \right\rangle \, .
\label{eq:eikonals}
\end{eqnarray}
\begin{figure}[t]
\centering
\includegraphics[scale=0.5,angle=0]{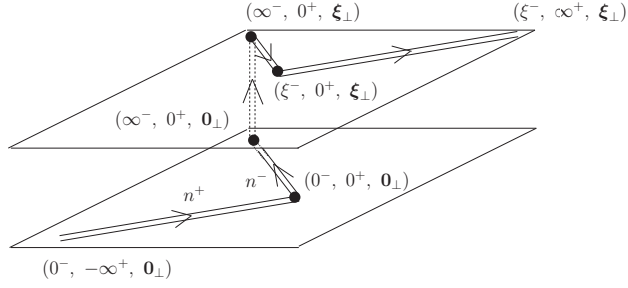}~~
\caption{Integration contour associated with the additional soft
         counter term.
\label{fig:contour}}
\end{figure}
The soft counter term $R$ has to be evaluated along the cusped
integration contour
($n_\mu^-$: minus light-cone vector)
\begin{equation}
\mathcal{C}_{\rm cusp}:\zeta_\mu
=
  \left\{
         [p_\mu^{+}s, - \infty < s < 0]
         \cup [n_\mu^-  s^{\prime},
         0 < s^{\prime} < \infty] \cup
         [ \mbox{\boldmath$l_\perp$} \tau , 0 < \tau < \infty ]
  \right\} \, ,
\label{eq:contour}
\end{equation}
depicted in Fig.\ \ref{fig:contour}.
The main characteristic of this contour is the jump in the
four-velocity $v_1 = p^+$ (parallel to the plus light-cone ray) at the
origin to $v_2 = n^-$ (parallel to the minus light-cone
ray), and creating an angle-dependence via $(v_1 \cdot v_2) = p^+$.
Hence, the contour $\mathcal{C}$ is cusped with an angle
$\chi \sim \ln p^+ = \ln (p \cdot n^-)$.
This way, we arrive at the following redefined expression for the TMD
PDF:
\begin{eqnarray}
f_{q/q}^{\rm mod}\left(x, \mbox{\boldmath$k_\perp$};\mu, \eta
                   \right)
\!\!\!\!\!\! && \!\!\! =
  \frac{1}{2}
  \int \frac{d\xi^- d^2\mbox{\boldmath$\xi_\perp$}}{2\pi (2\pi)^2}
  \exp\left(- i k^+ \xi^-
   + i \mbox{\boldmath$k_\perp$}
\cdot \mbox{\boldmath$\xi_\perp$}\right)
\nonumber \\
&& \times
  \Big\langle
              q(p) |\bar \psi (\xi^-, \mbox{\boldmath$\xi_\perp$})
              [\xi^-, \mbox{\boldmath$\xi_\perp$};
   \infty^-, \mbox{\boldmath$\xi_\perp$}]^\dagger
   \nonumber \\
&& \times
   [\infty^-, \mbox{\boldmath$\xi_\perp$};
   \infty^-, \mbox{\boldmath$\infty_\perp$}]^\dagger
   \gamma^+[\infty^-, \mbox{\boldmath$\infty_\perp$};
   \infty^-, \mbox{\boldmath$0_\perp$}] \nonumber \\
&& \times
   [\infty^-, \mbox{\boldmath$0_\perp$}; 0^-,\mbox{\boldmath$0_\perp$}]
   \psi (0^-,\mbox{\boldmath$0_\perp$}) |q(p)
   \Big\rangle \nonumber \\
&& \times
   \Big[ \Phi(p^+, n^- | 0^-, \mbox{\boldmath$0_\perp$})
   \Phi^\dagger (p^+, n^- | \xi^-, \mbox{\boldmath$\xi_\perp$})
   \Big] \, ,
\label{eq:final}
\end{eqnarray}
which is the core result of our analysis.

\section{Conclusions}
\label{sec:concl}

We have given a modified definition of the TMD PDF for the unpolarized
case which incorporates additional eikonal factors in order to
cancel an anomalous-dimension contribution, stemming from the
cusped-like junction of the gauge contours at light cone infinity in
the transverse direction.
These transverse gauge links are indispensable for the sake of full
gauge invariance in the light-cone gauge, but would yield without these
corrective eikonal factors to results that differ from those in
covariant gauges in line with factorization.
Integrating over the transverse momentum, our expression
(\ref{eq:final}) provides an integrated PDF, which obeys the DGLAP
evolution equation, without any dependence on pole prescriptions or
the gauge-contour obstructions.

\section*{Acknowledgements}
This work was supported in part by the Heisenberg-Landau Programme
(grant 2008), the Deutsche Forschungsgemeinschaft under contract
436RUS113/881/0, and the Russian Federation President's Grant
1450-2003-2.

\end{document}